\documentclass[cits]{PoS}

\title{High-energy emission of variable objects in the OMC--VAR catalogue}

\ShortTitle{High-E counterparts in OMC--VAR}

\author{\speaker{Julia Alfonso-Garz\'{o}n}\\
        Centro de Astrobiolog\'{\i}a (INTA-CSIC), Madrid, Spain\\
        E-mail: \email{julia@cab.inta-csic.es}}

\author{Albert Domingo\\
        Centro de Astrobiolog\'{\i}a (INTA-CSIC), Madrid, Spain\\
        E-mail: \email{albert@cab.inta-csic.es}}

\author{J. Miguel Mas-Hesse\\
        Centro de Astrobiolog\'{\i}a (INTA-CSIC), Madrid, Spain\\
        E-mail: \email{mm@cab.inta-csic.es}}

\abstract{OMC--VAR is the first catalogue of variable sources observed by the Optical Monitoring Camera (OMC) on board INTEGRAL. It includes photometry and variability data for more than 5000 sources of very different nature. In order to study the multi-wavelength similarities and differences of the engines powering AGN and X-ray binaries, especially those powered by a black hole, we have searched for counterparts in several high-energy and infrared catalogues, including the 4th IBIS/ISGRI soft gamma-ray survey catalogue, the XMM-Newton Serendipitous Source catalogue, the 2MASS All-Sky catalogue of Point Sources, the 2MASS Extended Source Catalog and the WISE All-Sky Data Release.
Preliminary multiwavelength results for the IBIS counterparts are  presented and discussed. 
}

\FullConference{"An INTEGRAL view of the high-energy sky (the first 10 years)"\\
9th INTEGRAL Workshop and celebration of the 10th anniversary of the launch,\\
		October 15-19, 2012\\
		Bibliotheque Nationale de France, Paris, France}

\begin{document}

\section{The first INTEGRAL--OMC catalogue of optically variable sources}
The Optical Monitoring Camera (OMC) on board the high-energy observatory INTEGRAL \cite{Winkler03} provides photometry in the Johnson V band of the same fields observed by the gamma-ray instruments \cite{MasHesse03}. With an aperture of 50 mm and a field of view of $5\,^{\circ}\times5\,^{\circ}$, OMC is able to detect optical sources brighter than V$\sim$18, from a previously selected catalogue. 
In the first version of OMC--VAR, the catalogue of variable sources observed by OMC \cite{Alfonso12}, we provide for each object the median of the visual magnitude, the magnitude at maximum and minimum brightness in the light curve during the observing window, the period when found, as well as the complete intrinsic and period-folded light curves, together with some ancillary data.

\begin{figure}[h]
\centering
\includegraphics[width=0.7\textwidth]{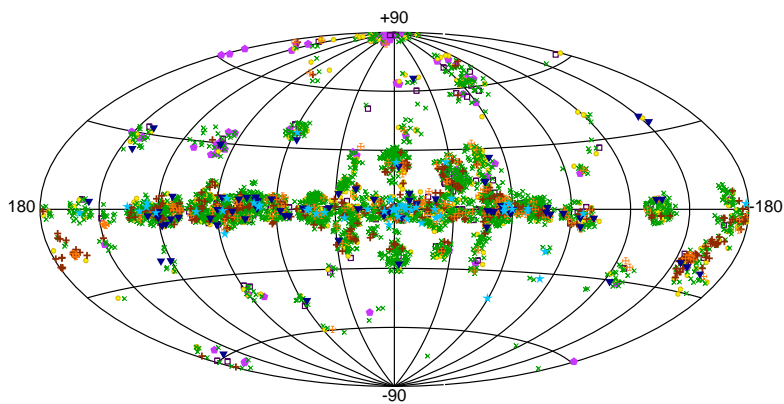}
\caption{Distribution in galactic coordinates of all the sources in the catalogue. The green crosses represent pulsating stars, the red filled points correspond to eclipsing binaries, the brown plusses are eruptive stars, the orange complex plusses represent rotating stars, the inverted dark blue filled triangles represent cataclysmic variables, the light blue filled stars are X-ray binaries, the yellow filled points correspond to objects simply classified as variable stars, the purple filled pentagons represent extragalactic objects and the empty purple squares are other types of objects.}
\label{fig1}
\end{figure}

There is a big variety of variable objects in this version of the catalogue. We show in Fig. \ref{fig1} the distribution in galactic coordinates of the different kinds of objects, that are located in the regions where INTEGRAL has devoted more observing time in its first 10 years of operations. 
Illustrative examples of the charts provided in the catalogue for different sources are shown in Fig. \ref{fig2}. In all cases we include the DSS red image of the field of view (top left panel -- red circles represent the OMC photometric aperture), and the non-folded OMC V-band light curve (top right). In the bottom panels we show the folded light curves (with OMC and VSX periods, when available), or the histograms of observed magnitudes. 

Similar charts and the complete light curves in machine readable format for all the sources contained in the OMC--VAR catalogue can be retrieved from \textit{http://sdc.cab.inta-csic.es/omc/} and from the CDS  via \mbox{\textit{http://cdsweb.u-strasbg.fr/cgi-bin/qcat?J/A+A/548/A79/}}. The time coordinate in the light curves is the Barycentric INTEGRAL Julian Date expressed in Barycentric Dynamical Time (TDB). To convert this time to Barycentric Julian Date expressed in TDB, you have to add 2,451,544.5 d.

\begin{figure}
\centering
\includegraphics[width=0.65\textwidth, angle=-90]{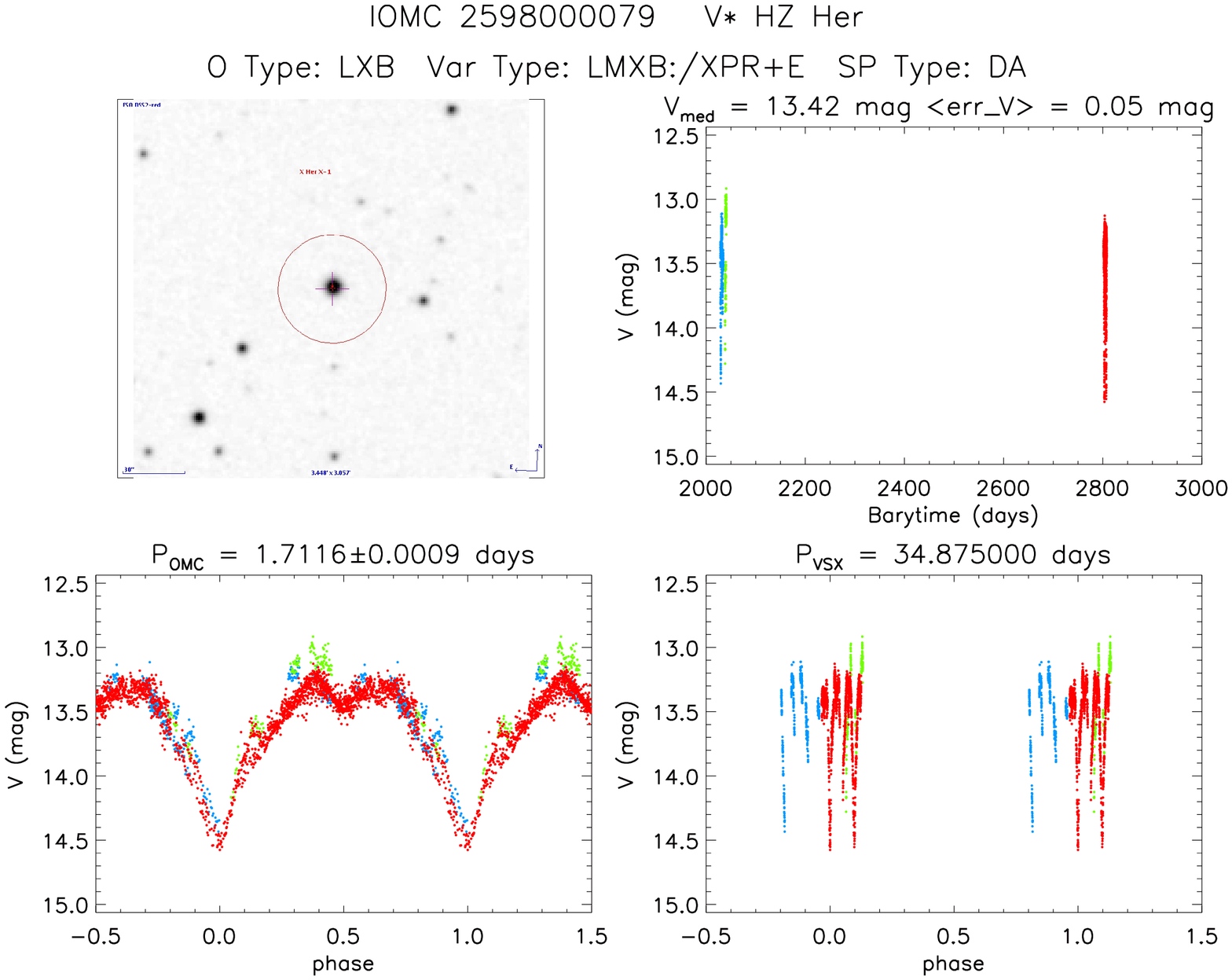}
\includegraphics[width=0.65\textwidth, angle=-90]{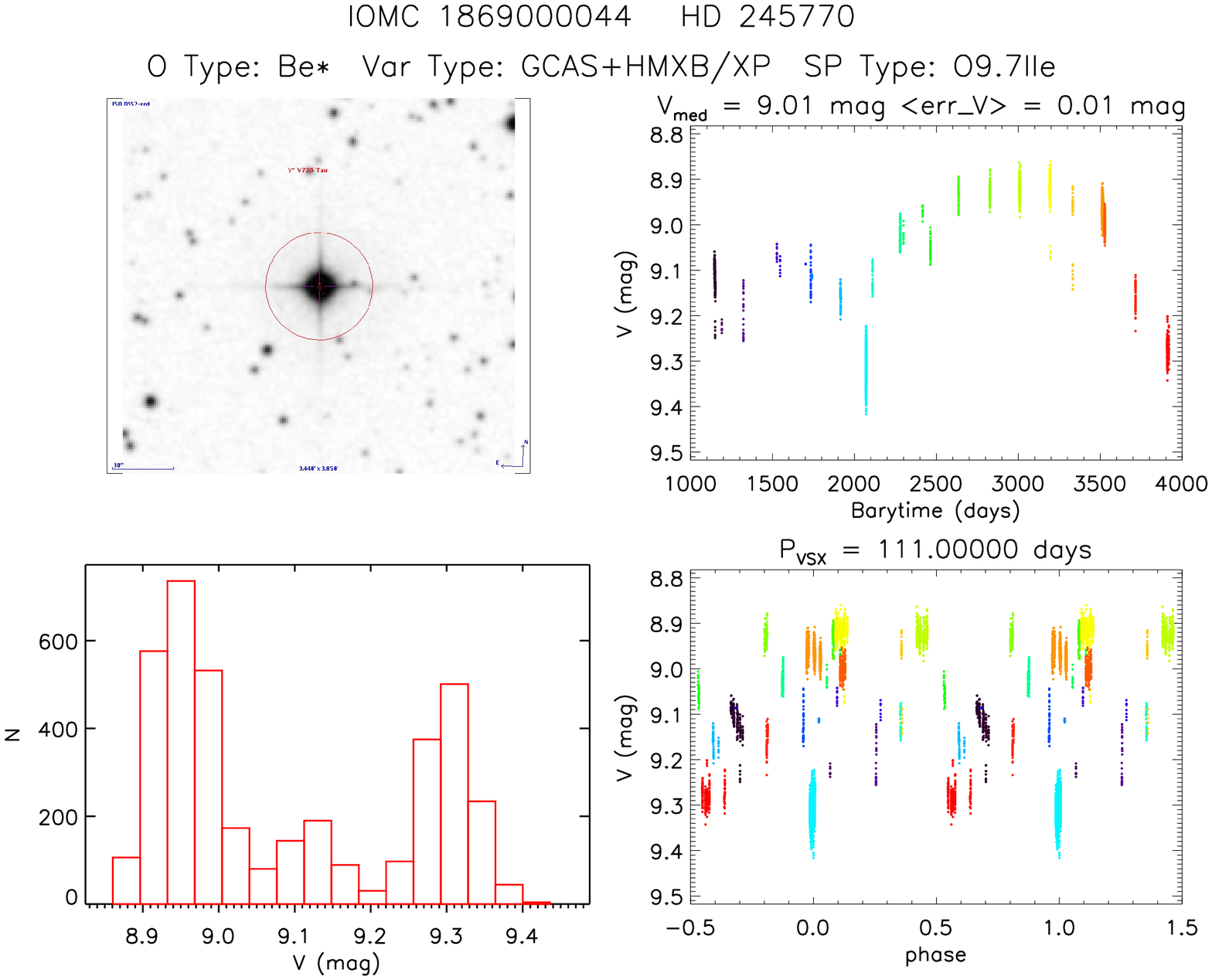}
\caption{Examples of the information provided in the catalogue. \textbf{Top:} For the LMXB Her X-1 the VSX period corresponds to the 35-day cycle in the X-ray observations, while the OMC optical light curve is dominated by the eclipses.   \textbf{Bottom:} In the light curve of HMXB 1A 0535+262 we see a long term variation during the nearly 3000 days of monitoring. We can not derive a period, and the value given by the VSX does not seem to fold properly the light curve.}
\label{fig2}
\end{figure}

\addtocounter{figure}{-1}
\begin{figure}[t]
\centering
\includegraphics[width=0.65\textwidth, angle=-90]{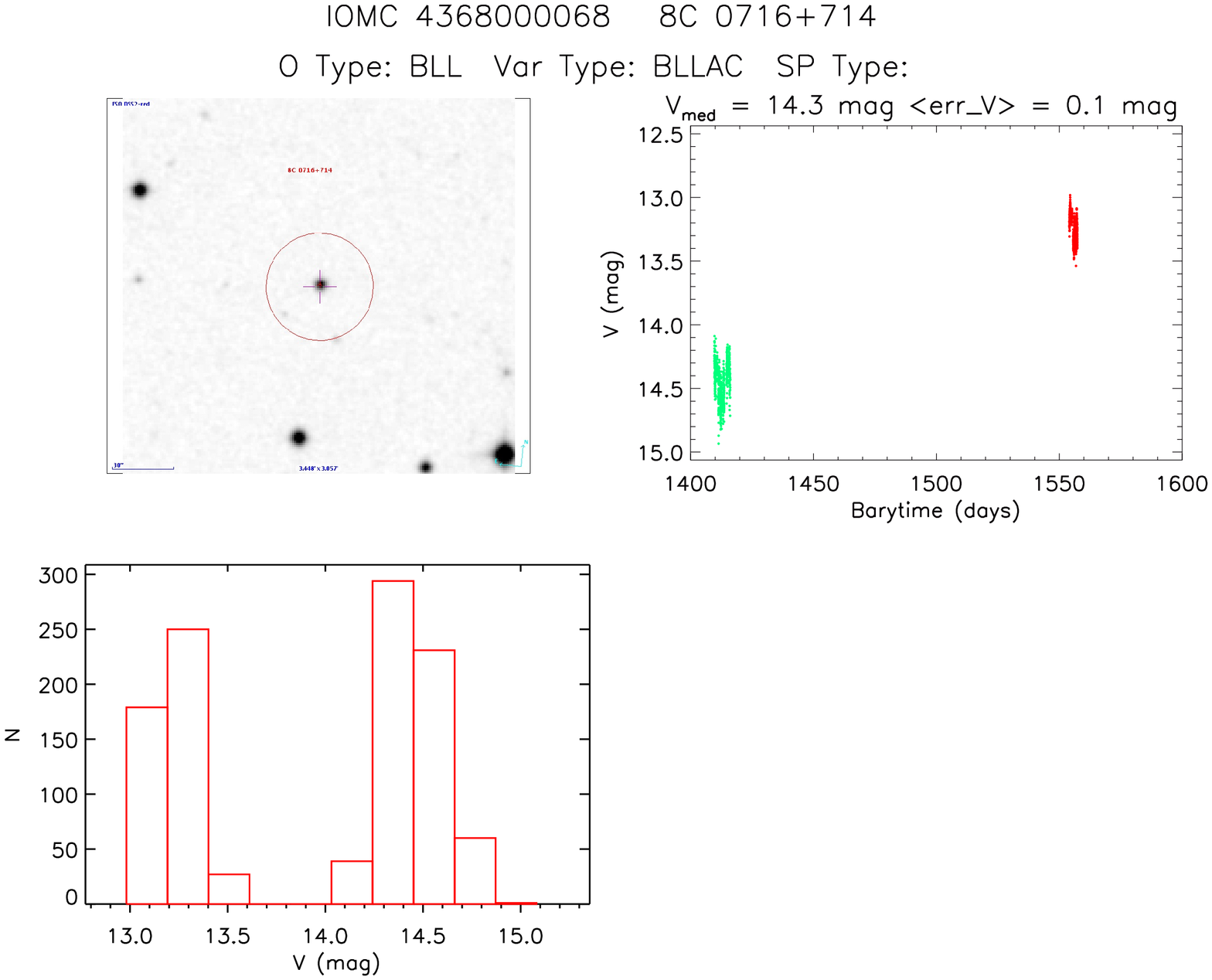}
\caption{\textit{Cont.} The OMC light curve of the blazar QSO B0716+714 shows two different states, separated by around 150 days. }
\label{fig2-2}
\end{figure}

\section{Multi-wavelength analysis}
\begin{figure}
\centering
\includegraphics[width=0.5\textwidth]{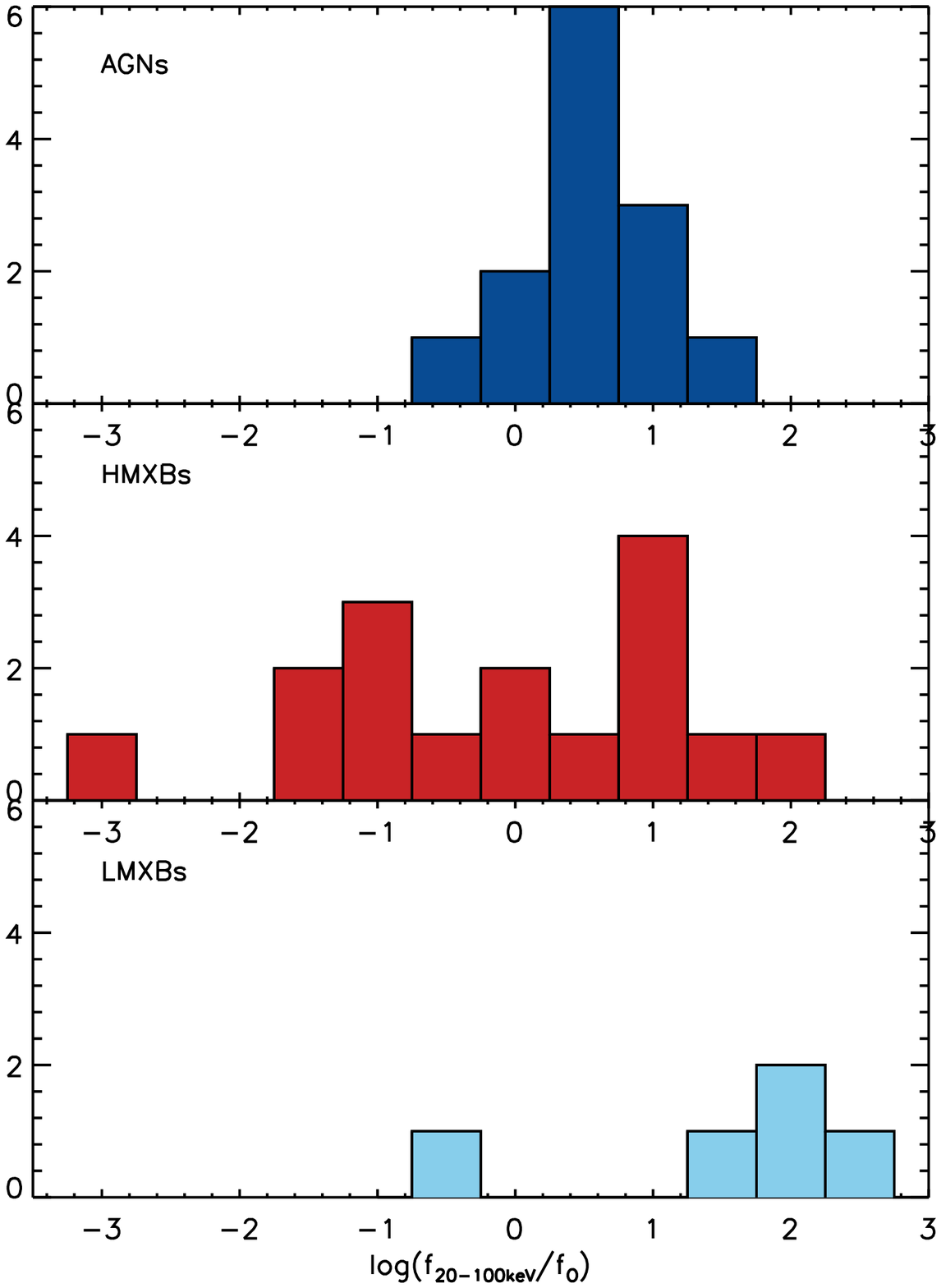}
\caption{Histograms of $F_{20-100keV}$ / $F_{OMC}$. \textbf{Top:} AGN, \textbf{middle:} HMXBs and \textbf{bottom:} LMXBs.}
\label{fig3}
\end{figure}

Our sample consists of the 49 IBIS high-energy counterparts found after cross-correlating with our catalogue. These data have been complemented with 2MASS and WISE photometry. We have calculated the fluxes in the different bands using the relations log($F_{OMC}$) = - V$/$2.5 - 5.49  from \cite{Bessell90}, log($F_{NIR}$) = - $K_{s}$ / 2.5 - 6.95, taking the in-band zero magnitude flux of the $K_{s}$ band from \cite{Cohen03}, and log($F_{MIR}$) = - W3 / 2.5 - 10.45 from \cite{Jarrett11}. Fluxes are given always in erg s$^{-1}$ cm$^{-2}$. We show in Fig. \ref{fig3} the histograms of $F_{20-100keV}$ / $F_{OMC}$ for the three subsamples we have considered in our analysis: AGN, HMXBs and LMXBs. As already discussed in \cite{Beckmann09}, $F_{20-100keV}$ / $F_{OMC}$  peaks in AGN around 0.6, showing a rather narrow distribution. While the average value for HMXBs is similar, they show a much broader distribution, indicating the apparent lack of correlation between the high-energy engine and the optical luminosity of the companion star.  In LMXB there is a clear lack of  optical flux, as expected from the low mass of their  stellar companion. 

\begin{figure}
\centering
\includegraphics[width=0.6\textwidth]{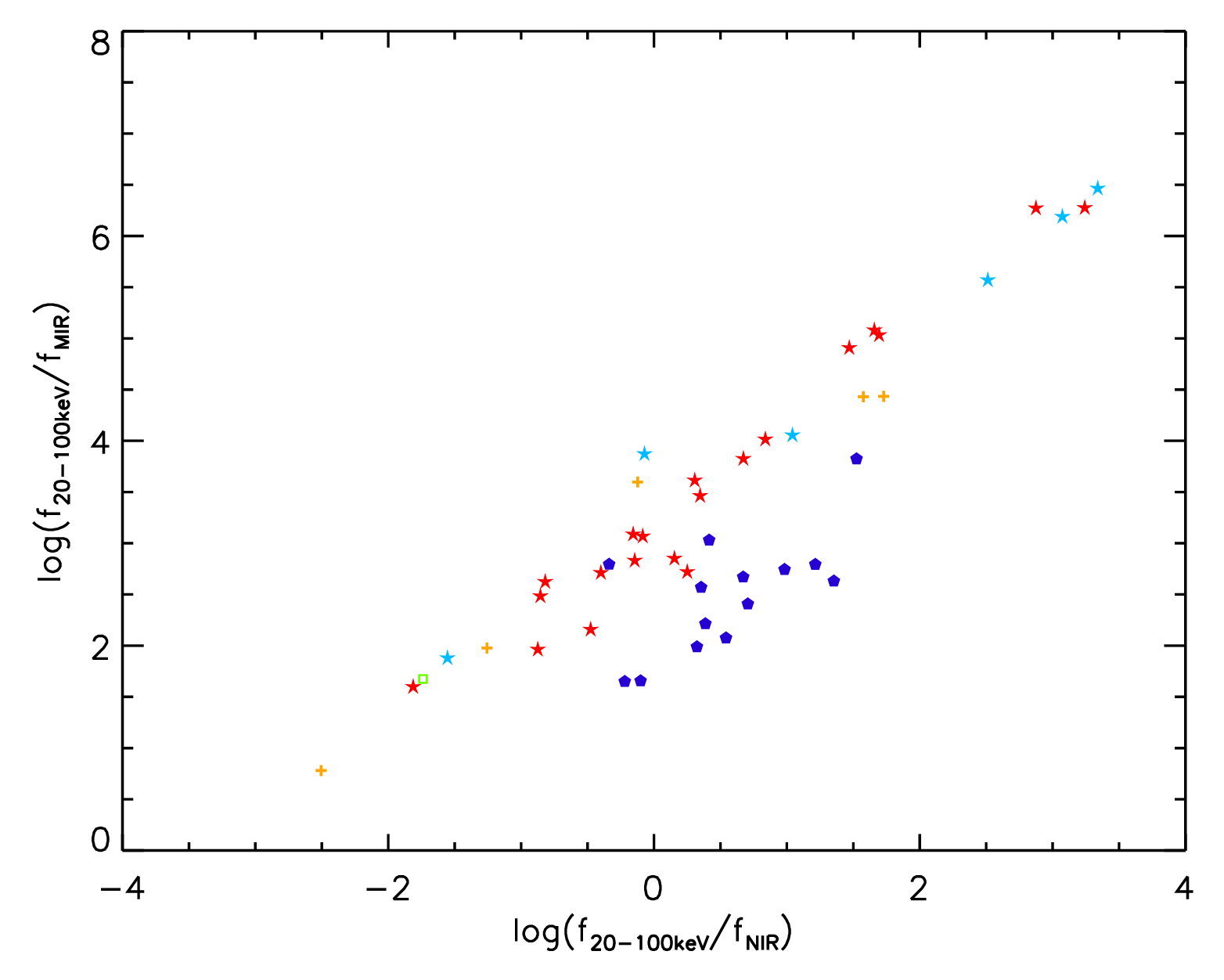}
\caption{$F_{20-100keV}$ / $F_{MIR}$ vs. $F_{20-100keV}$ / $F_{NIR}$ for the IBIS counterparts in OMC--VAR. Blue pentagons represent AGN, red stars are HMXBs, light blue stars are LMXBs and orange crosses are CVs.}
\label{fig4}
\end{figure}

\begin{figure}
\includegraphics[width=.5\textwidth]{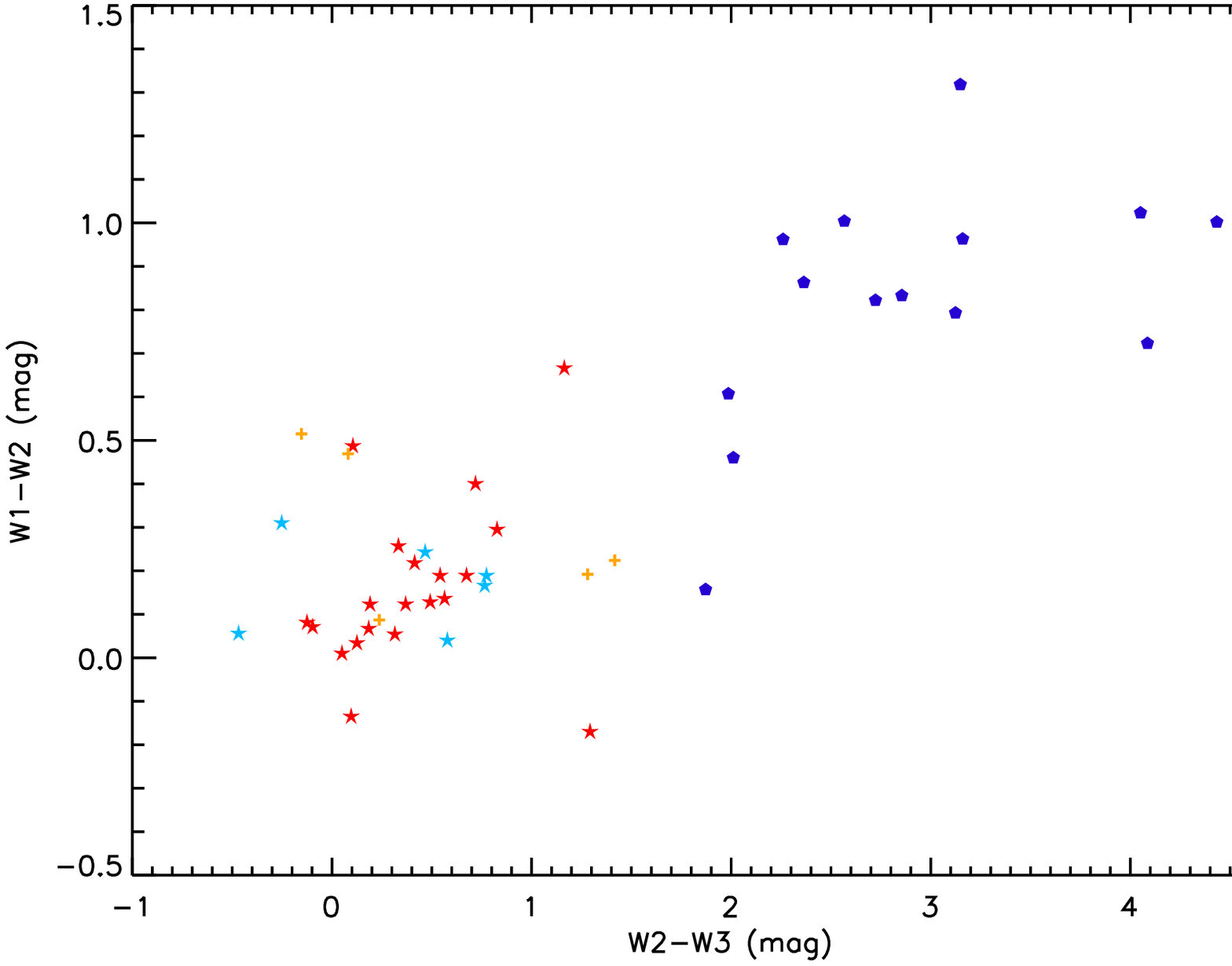}
\includegraphics[width=.5\textwidth]{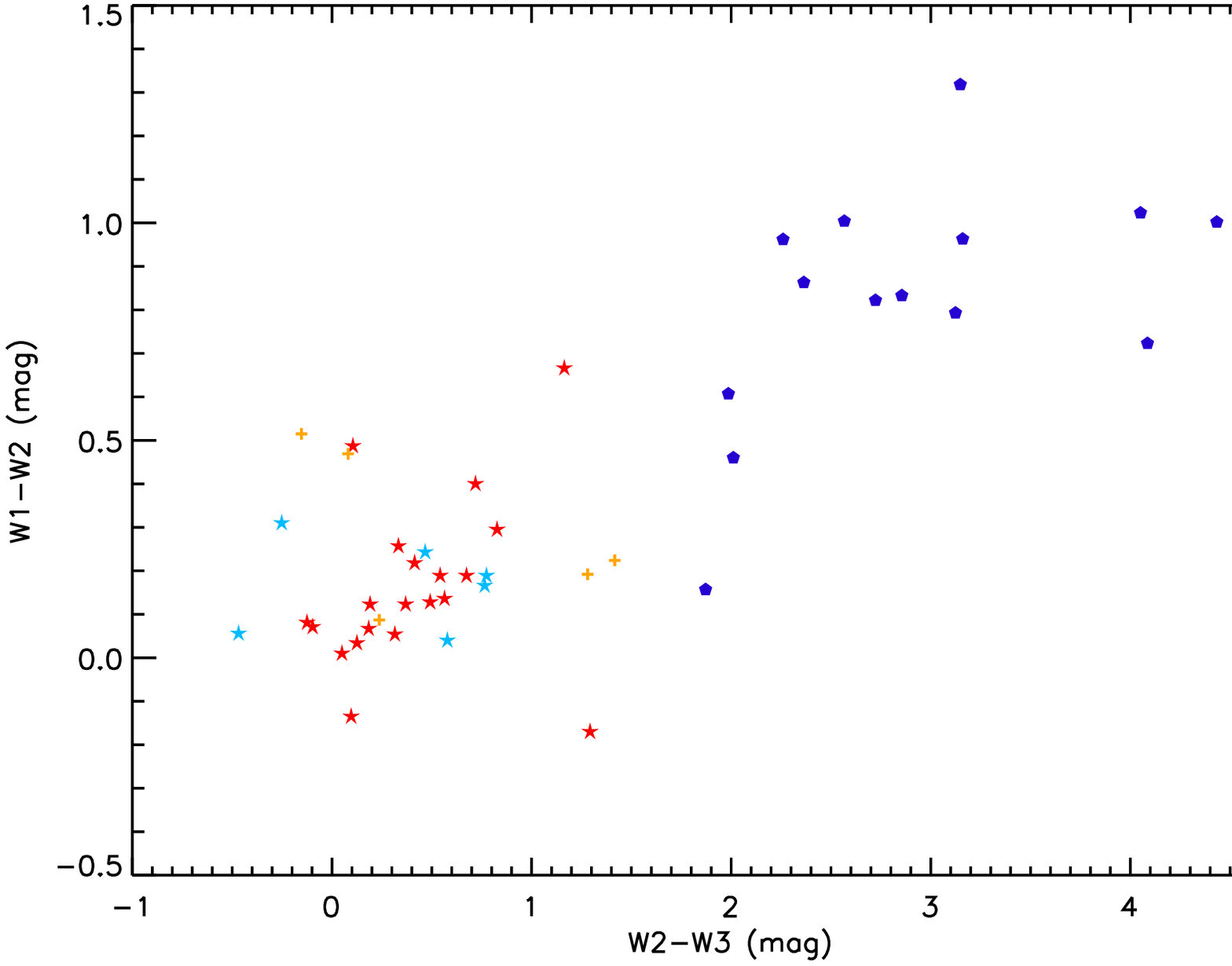}
\caption{\textbf{Left:} Near-infrared colours from 2MASS (J = 1.235 $\mu$m, H = 1.662 $\mu$m, $K_{s}$ = 2.159 $\mu$m). \textbf{Right:} Mid-infrared colours from WISE (W1 = 3.4 $\mu$m, W2 = 4.6 $\mu$m, W3 = 12 $\mu$m, W4 = 22 $\mu$m). Symbols as in Fig. 4.}
\label{fig5}
\end{figure}

The $F_{20-100keV}$ / $F_{MIR}$ vs. $F_{20-100keV}$ / $F_{NIR}$ diagram shown in  Fig. \ref{fig4}  clearly separates the AGN population from the HMXBs and LMXBs. This is due to the emission in the mid IR associated to the torus surrounding the core engine in AGN.  The effect of the torus becomes also clearly visible in Fig. \ref{fig5}: while both populations (AGN and binaries) overlap when their infrared colours from 2MASS are considered, the WISE mid-infrared colours (MIR) colours are dominated by the emission from the torus separating the AGN from the binaries population. The MIR emission from the accretion disc is probably hidden in AGN within the strong emission by the torus, originated by the reprocessing of the strong UV-X ray flux originated by the central engine. This reprocessed luminosity is responsible for the systematic low values of $F_{20-100keV}$ / $F_{MIR}$ in Fig. \ref{fig4}. 

The expected intrinsic similarities between the engines powering AGN and BH XBs (after scaling!) seem therefore affected by their different surroundings, producing clear multi-wavelength differences. We are now analysing their OMC and IBIS light curves, looking for systematic intrinsic similarities and differences in the accretion processes onto massive black holes in AGN and stellar mass black holes, or compact objects, in binaries of different types. The results will be published elsewhere in the near future. 

\acknowledgments
The development, operation and exploitation of OMC have been funded by Spanish MICINN
under grants AYA2008-03467 and AYA2011-24780 and previous ones.  
INTEGRAL is an ESA project funded by ESA member states (especially the PI countries:
Denmark, France, Germany, Italy, Spain, Switzerland), Czech Republic, Poland, and with the
participation of Russia and the USA. OMC was also funded by Ireland, United Kingdom, Belgium and the Czech Republic.
This research has made use of data from the OMC Archive at CAB (INTA-CSIC), pre-processed
by ISDC, and of the SIMBAD database, operated at CDS, Strasbourg, France.

\end{document}